\documentclass[conference]{IEEEtran} 
\usepackage{cite}

\usepackage{xcolor}
\usepackage{fancyhdr}
\ifCLASSINFOpdf
  \usepackage[pdftex]{graphicx}
\else

\fi

\usepackage{amsmath}
\usepackage{url}

\usepackage{tikz}
\usepackage{textcomp}
\usepackage{hyperref}
\usepackage{lipsum}

\newcommand\copyrighttext{%
  \footnotesize \textcopyright 2020 Crown ~~\textcolor{blue}{
}~\\ \underline{Citation:}{ P. Hu, "IoT-based Contact Tracing Systems for Infectious Diseases: Architecture and Analysis," GLOBECOM 2020 - 2020 IEEE Global Communications Conference, Taipei, Taiwan, 2020, pp. 1-7, doi: 10.1109/GLOBECOM42002.2020.9347957.
}
}
\newcommand\copyrightnotice{%
\begin{tikzpicture}[remember picture,overlay]
\node[anchor=south,yshift=10pt] at (current page.south) {\fbox{\parbox{\dimexpr\textwidth-\fboxsep-\fboxrule\relax}{\copyrighttext}}};
\end{tikzpicture}%
}

\begin{document}
\bstctlcite{IEEEexample:BSTcontrol}

\title{IoT-based Contact Tracing Systems for Infectious Diseases: Architecture and Analysis}

\author{\IEEEauthorblockN{Peng Hu}
\IEEEauthorblockA{\textit{Digital Technologies Research Center} \\
\textit{National Research Council of Canada}\\
Waterloo, Canada \\
Peng.Hu@nrc-cnrc.gc.ca}}

\markboth{Journal of \LaTeX\ Class Files,~Vol.~00, No.~00, 00~2020}%
{Shell \MakeLowercase{\textit{et al.}}: Bare Demo of IEEEtran.cls for IEEE Journals}

\IEEEoverridecommandlockouts
\IEEEpubid{\makebox[\columnwidth]{
 \hfill} \hspace{\columnsep}\makebox[\columnwidth]{ }}

\maketitle
\copyrightnotice

\begin{abstract}
The recent COVID-19 pandemic has become a major threat to human health and well-being. Non-pharmaceutical interventions such as contact tracing solutions are important to contain the spreads of COVID-19-like infectious diseases. However, current contact tracing solutions are fragmented with limited use of sensing technologies and centered on monitoring the interactions between individuals without an analytical framework for evaluating effectiveness. Therefore, we need to first explore generic architecture for contact tracing in the context of today's Internet of Things (IoT) technologies based on a broad range of applicable sensors. A new architecture for IoT based solutions to contact tracing is proposed and its overall effectiveness for disease containment is analyzed based on the traditional epidemiological models with the simulation results. The proposed work aims to provide a framework for assisting future designs and evaluation of IoT-based contact tracing solutions and to enable data-driven collective efforts on combating current and future infectious diseases. 
\end{abstract}


%
\IEEEpeerreviewmaketitle

\section{Introduction}

The recent COVID-19 pandemic has become a major threat to human health and well-being. With the absence of effective vaccines and pharmaceutical treatments, non-pharmaceutical interventions (NPIs) \cite{Ferguson2020} through reducing contact rates and contact tracing of suspected exposure to an infectious disease are essential to containing emerging epidemics \cite{Peak2017}. The use of Internet-based systems has shown its importance in pandemics \cite{Salathe13}, but there is still much room for improvement for the current non-pharmaceutical interventions based on the traditional reporting model. With the advanced Internet of Things (IoT) and artificial intelligence (AI) technologies, autonomous contact tracing for infectious diseases can be enabled. It is the right time to examine how IoT-based contact tracing system (IoT-CTS) can help with the disease containment.

Current implementations on contact tracing are centered on user interactions between individuals. These so-called ``peer-to-peer'' (P2P) contact tracing solutions need to detect locations or proximity of users, which can be done in various means. Positioning or navigation devices such as GPS, Bluetooth \cite{Farrahi2014,Dong19, Apple2020}, mobile networks \cite{TheEconomist20}, and thermal cameras can be used to detect the location traces or proximity of users automatically. There are solutions using other sensors such as magnetometers \cite{Jeong19} for inferring the proximity between users. Contact tracing can be also performed through manual scanning of bar codes or QR codes on mobile devices \cite{YashengHuang} and through the analysis with the activity traces on social network platforms. However, an IoT-CTS can involve a broad use of various sensing technologies that have not been addressed systematically from its design to evaluation. 

With different sensing and connectivity options of an IoT system, we may easily design a solution for contact tracing. However, we need to understand two basic problems: what we can do with an IoT-CTS for contact tracing and how to evaluate it in the design phase? The first problem relates to the design of a generic IoT-CTS, where the context of system architecture and the mobility in geographical regions need to be considered. In that context, the system capabilities, characteristics, use cases, and deployment options can be clearly discussed. To address the second problem, we need to see how the IoT-CTS can affect the classical epidemiological models. 



In this paper, we will address the essential topics for the IoT-CTS designs from the perspective of key elements, architecture, and analytical methods. The key contributions made are summarized as follows:
\begin{itemize}
    \item Key elements including sensing technologies, sensing data and its usage for an IoT-CTS are systematically discussed. 
    \item A generic architecture for IoT-CTS solutions is proposed where the protocol structure model, key entities and operational models are discussed.
    \item Analysis of the proposed solution using the epidemiological model is examined against the classical SIR model and is shown effective through the computer simulations in typical scenarios.
\end{itemize}

The rest of the paper is structured as follows. Section II presents the related work. Section III discusses the proposed system architecture. Section IV discusses the general analysis methods and performance evaluation. Section V presents conclusive remarks and future work.

\section{Related Work}
The latest works have been focused on multi-disciplinary research on contact tracing in interacting networks. Farrahi \textit{et al.} \cite{Farrahi2014} studied how to reconstruct physical interactions from the mobile phone communication logs on calls, short messages, and Bluetooth interactions, where the physical interaction networks and communication networks are modelled as a dual network setting where the latter can be viewed as ``proxies'' of the former. Alvarez Zuzek \textit{et al.} \cite{Farrahi2014} studied the isolation effect based on the SIR model following a two-layer network, with one layer in the work environment and the other layer in the social environment. Salath{\'{e}} \textit{et al.} \cite{Salathe13} indicated the use of Internet-based systems for surveillance such as the Internet, mobile phones, and social media  can provide ``important early epidemic intelligence for the 2003 SARS and 2009 H1N1 influenza pandemic''. 

An IoT-CTS relies on sensing technologies, where multiple kinds of them can be used. On the one hand, an automatic P2P contract tracing solution with proximity sensing can be realized through various sensing mechanisms such as wireless, ultrasonic, inductive, and capacitive sensing. The proximity sensing mechanisms may result in different precision performance at a distance. The wireless sensing technologies can use RFID based solutions such as near-field communication (NFC) in an active or passive fashion or regular wireless personal-area networking (WPAN) technologies such as IEEE 802.15.4 and Bluetooth. The encoded signals sent through electromagnetic (EM) or acoustic waves can contain messages \cite{Hanspach2013}. On the other hand, a manual P2P contract tracing solution can be done through multiple digital ways such as QR codes \cite{YashengHuang, TheEconomist20}. 

The use of these sensing mechanisms would require the support of hardware and software on user devices and there are a few sensing technologies that have been utilized in recent implementations. A solution using magnetometers on smartphones based on magnetic induction has been proposed in \cite{Jeong19}, where magnetometer readings of two smartphone users are used to determine close proximity. A P2P contact tracing app ``TrackCOVID'' was developed \cite{Yasaka20} with privacy preservation. The spatial-proximity information is also shown to be able to infer the flu based on Wi-Fi and Bluetooth scanning with the PocketCare mobile app \cite{Dong19}. Apple and Google announced the plan for contact tracing apps called ``Privacy-Preserving Contact Tracing'' (https://www.apple.com/covid19/contacttracing/), where the new ``Contact Detection Service'' for Bluetooth Low Energy (BLE) \cite{Apple2020} will be used as the underlying sensing technology. From \cite{Apple2020}, the ``contact tracing makes it possible to combat the spread of the COVID-19 virus by alerting participants of possible exposure to someone who they have recently been in contact with, and who has subsequently been positively diagnosed as having the virus.'' The PrivateKit mobile app (http://privatekit.mit.edu/) developed by MIT uses GPS and Bluetooth trails for contact tracing with privacy. In addition, there have been increasing global efforts on making available data for health research. For example, in March 2020, the Allen Institute for AI with other research groups have prepared the COVID-19 Open Research Dataset (CORD-19) with COVID-19 related full-text articles. In early April 2020, Facebook released the new data tools for its Data for Good program \cite{Jin2020} to combat COVID-19.

The current applications mainly focus on interactions between users and the use of sensing technologies for non-pharmaceutical contact tracing is in the early stage and many efforts in hardware and software have yet to be done. The discussions on key elements and architecture for an IoT-CTS using various sensing technologies for tracing all essential events for infectious disease are lacking in the literature, which will be addressed in the paper.

\section{Key Elements and System Architecture}
One characteristic of the IoT-CTS solutions compared to the traditional contact tracing solutions is the use of sensing technologies with Internet-based connectivity. Let us first look at the key elements of an IoT-CTS in the following subsections. 

\subsection{Sensing Technologies for IoT-CTS Solutions}
Sensing capabilities are essential for IoT-CTS solutions and we need to see what and how different types of sensors can be used for a COVID-19-like infectious disease based on its transmission methods. Although COVID-19 is still under research, current information from the public health authorities \cite{HC2020,CDC2020} indicates that COVID-19 spreads from person to person through respiratory droplets and from contaminated surfaces/objects (which are collectively referred as ``objects'' later) with different viability on various surfaces \cite{VanDoremalen2020}. The contact tracing, therefore, needs to address person-to-person and person-to-object/object-to-person transmissions. The person-to-person transmission is for tracking peer-to-peer contacts between a person in Susceptible (S) or Infectious (I) states and a healthy person. The person-to-object tracing is to trace a person in S/I states who may contaminate objects. The object-to-person is for any healthy person who may be infected by the contaminated objects. In this case, we can enable the contact tracing in a broader context such as when a person enters a building where sensing capability for person-to-object or object-to-person is deployed on the premises. For clarity, we will refer person-to-object or object-to-person as object-to-person later in the rest of the paper.

Detecting the exposure to infectious diseases through the person-to-person and object-to-person transmissions require sensing technologies. The possible sensing technologies are listed in Table \ref{Tbl:sensingspread} in alignment with the known methods of COVID-19 transmissions. The example sensor devices in Table \ref{Tbl:sensingspread} list the typical sensing devices, where ``mobile'' here generally refers to mobile devices with the presence of mobile networks, as it is often used to provide communication traces where possible social media traces from installed apps may also be utilized.

\begin{table}[htbp]
\caption{Types of Sensing technologies for various transmission modes}
\label{Tbl:sensingspread}
\centering
\begin{tabular}{|p{0.1\textwidth}|p{0.13\textwidth}|p{0.18\textwidth}|}
\hline
\textbf{Transmission mode} & \textbf{Sensing Type} & \textbf{Example Sensor Devices} \\ 
\hline
{Person-to-person} & Location-based & GPS, mobile \cite{Yasaka20, Farrahi2014, Dong19}, ultrasonic, Bluetooth\cite{Farrahi2014, Dong19}, magnetometer \cite{Jeong19} \\
 { }   & Computer vision & Camera, QR code \cite{TheEconomist20} \\
\hline
{Contaminated surfaces/objects (Object-to-person)} & Touch & Inductive/capacitive, RFID\cite{Hu2020} \\
 { }   & Distance, proximity & Ultrasonic, Bluetooth\cite{Farrahi2014, Dong19}, RFID\cite{Hu2020} \\
\hline
\end{tabular}
\end{table}

Generally speaking, these sensors shown in Table \ref{Tbl:sensingspread} can be operated in different modes, such as \textit{passive mode}, \textit{proactive mode}, and \textit{hybrid mode}. The passive mode means sensors automatically collect tracing-related data. The proactive mode means sensors collect tracing data with users' full control. For example, using the QR code in Table \ref{Tbl:sensingspread} would fall in the proactive mode where active user participation is required. The hybrid mode means sensors can collect tracing data partially in the proactive mode and partially in the passive mode. In the hybrid mode, a user may configure which sensors can collect data without active interventions and which sensors can collect data under the user's control and permission. The sensing devices may take various form factors such as a wearable device. Although the privacy considerations can be addressed in these modes, the in-depth discussions are out of the scope of this paper.

\subsection{Sensing Data}
Another important aspect of IoT-CTS solutions is the sensing data, where its essential features of tracking data should have the \textit{user identifier}, \textit{location}, and \textit{timestamp}. The identifier feature functions as the unique index of a user or an object which can be anonymized. The location feature is the coordinate of a geographic coordinate system, where the locations recorded in different coordinate systems should be able to be convertible between each other. Another useful feature can be the S/I/R state of a user, which can be used to log the basic states in the SIR model and extra attributes.

\begin{figure}[ht]
\centering
\includegraphics[width=\linewidth]{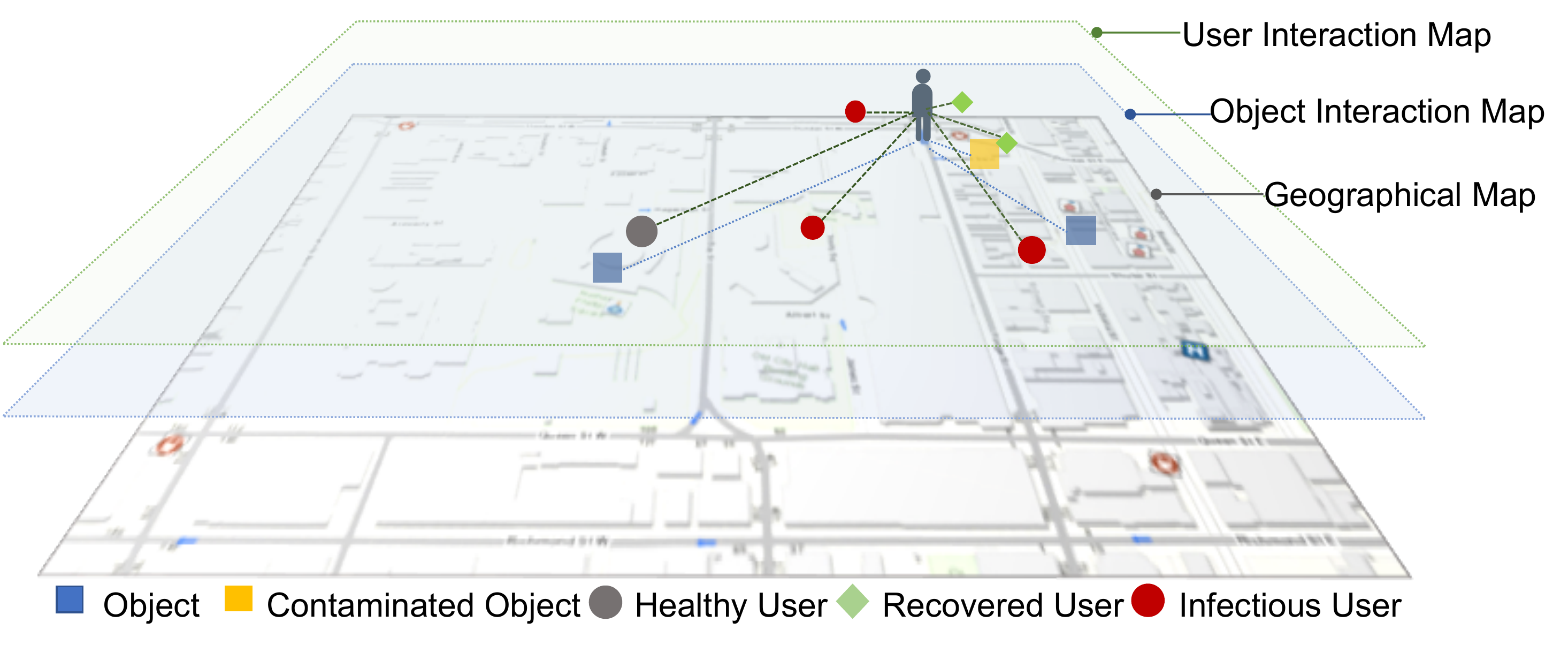}
\caption{An illustration visualizing the spatial-temporal sensing data collected, where a user interacted with other users and objects in different states are shown in objects interaction and user interaction maps. 
}
\label{tor_example}
\end{figure}

With these features, a spatial-temporal data of each user is available, which not only allows us to track the probability of getting infected and able to examine the transmission patterns over time. Furthermore, the granularity of contract tracing is indeed determined by the application requirements, which can be done with different deployment options of the sensing technologies. For example, if we make possible contact tracking in public transport, schools, and hospitals (as illustrated in Fig. \ref{tor_example}), we are able to track contact activities in these buildings, public transport facilities, and on public transport vehicles. In this way, if there is an infection risk occurred in a public facility at some time, users can be notified. The tracing data in other places can be available to a user using personal sensing devices (such as sensing devices on a mobile device) while the data usage is in compliance with a privacy setting or policy.

In the context of an IoT-CTS, we can assume a user is equipped with a sensing device with sensors that can meet the logging of the essential features of tracing data. The places that equipped with sensing devices for automatic contact tracing can log extra data.

To put the aforementioned discussions into a working IoT-CTS, we need to look at the architectural elements of it.

\subsection{Protocol stack model}
\begin{figure}[ht]
\centering
\includegraphics[width=0.8\linewidth]{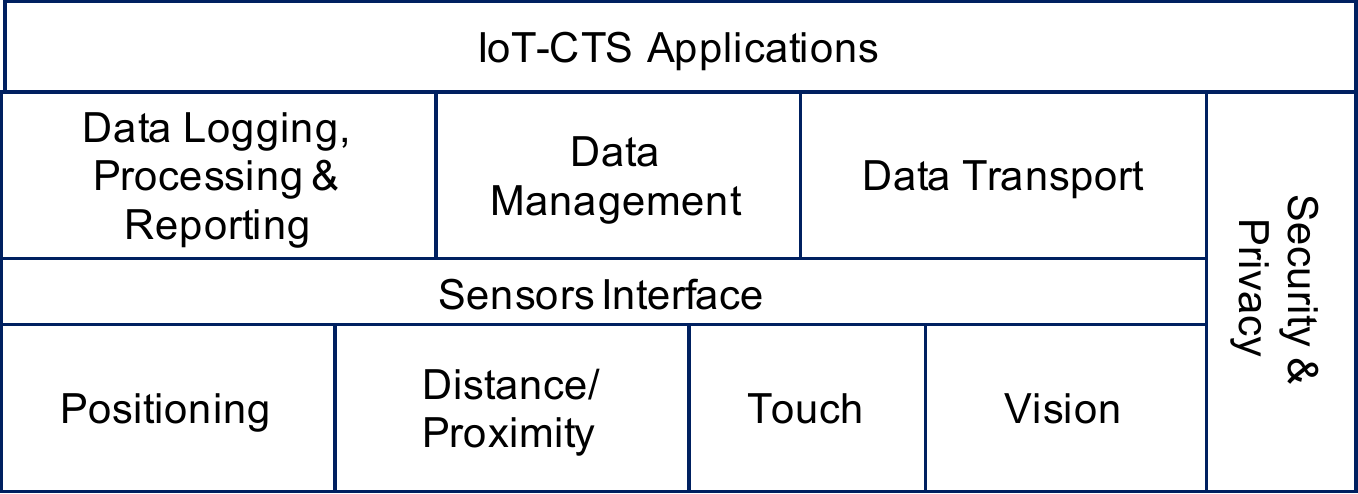}
\caption{Protocol stack model with key modules in an IoT-CTS}
\label{fig:block_diagram}
\end{figure}

A generic protocol stack model of an IoT-CTS is shown in Fig. \ref{fig:block_diagram} with the key modules including sensor components, sensors interface, data transport, data management, data logging, processing and reporting as well as security {\&} privacy. The sensor components can include a combination of sensors for positioning, proximity, touch, vision, etc., where sensors interface needs to be designed to configure and adapt the sensors hardware. The module of data logging, processing {\&} reporting handles the monitoring and processing of the sensing data and reporting mechanisms, where the data management module handles the data storage, retrieval, and configurations, which can be in compliance with the privacy settings through the security {\&} privacy module. The data transport module is responsible for transmitting the data to and interacting with an external endpoint securely. 

\subsection{Architectural Entities}
The proposed architecture for an IoT-CTS includes three entities: user endpoint (UE), facility endpoint (FE), and object endpoint (OE). An FE is an endpoint at a remote facility that can communicate with a UE or an OE. An FE can be a public health authority or a data center which can handle the data securely and be compliant with privacy policies and regulations. A UE is an endpoint for a sensing system equipped sensors hardware and software used by a user following the structure seen in Fig. \ref{fig:block_diagram}. Each user has a UE which can communicate with another UE on another user. An OE is an endpoint if a sensing system for an object or object surface, which can detect and monitor the touch events with sensors.

\begin{figure}[ht]
\centering
\includegraphics[width=\linewidth]{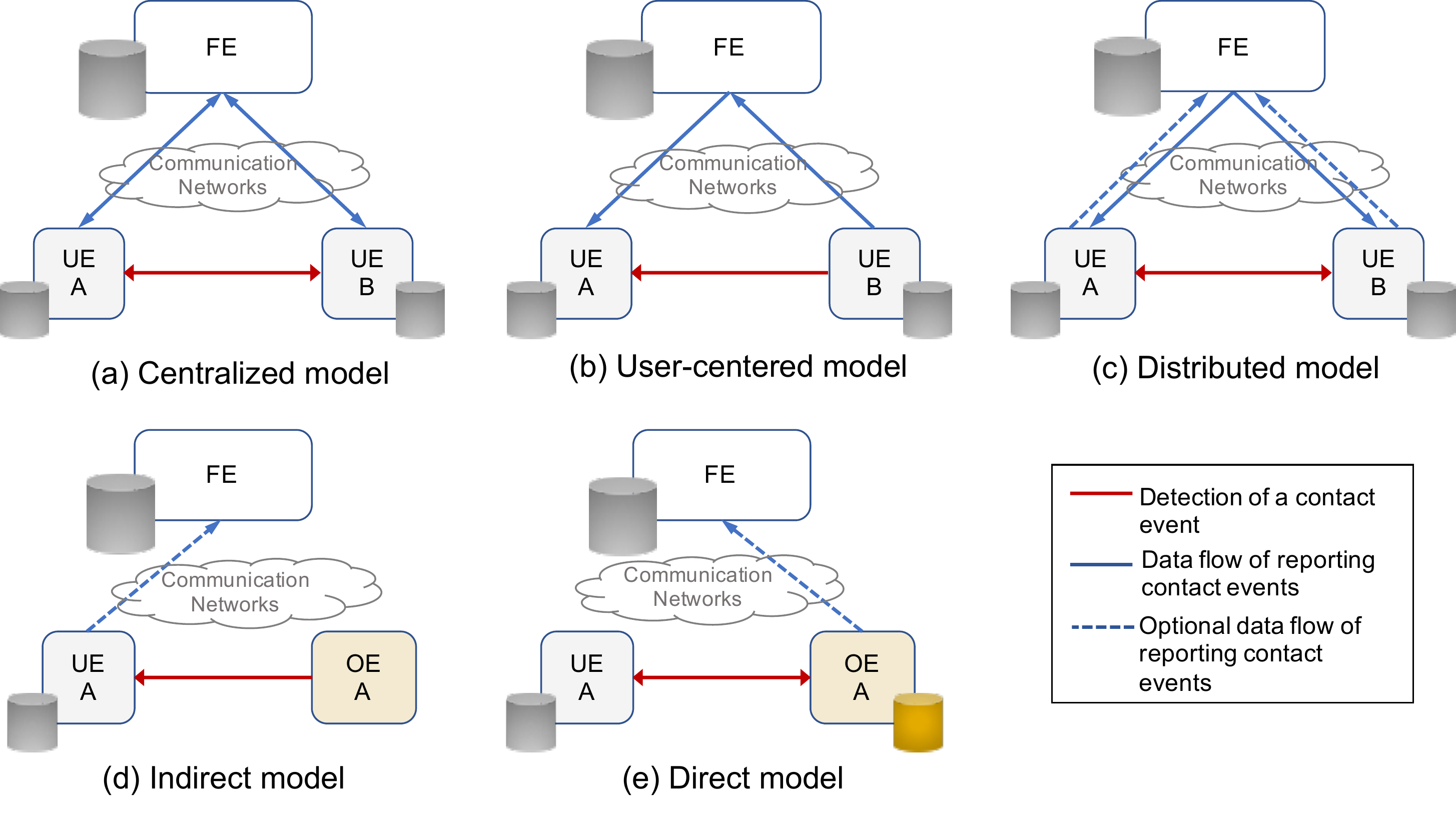}
\caption{Models of interactions between UEs, FEs, and OEs.}
\label{fig:modes}
\end{figure}

In Fig. \ref{fig:modes}, basic models of interactions between the basic IoT-CTS entities are shown, where the models based on the close proximity detected between UEs are the centralized model, user-centered model, and distributed model. In the centralized model, UE A and UE B can detect the close proximity between each other and report their own event data to an FE through communication networks. In this model, the FE has the full picture of the information, which can push data (such as notification of infectious risks) to the UEs and can enable UEs to retrieve data with authorization. The user-centered model allows a UE to be able to detect and record the contact events with other UEs and allows the UE to retrieve information from an FE. In the distributed model, UEs are able to exchange data between each other and to assess a user's infection risks locally but it can retrieve information from an FE and optionally transmit data to it. 

The models between a UE and an OE is shown in Fig. \ref{fig:modes} (d) and (e), where the indirect model means the contact event data between a user and an object is stored through the UE and the UE can optionally transmit data to the FE. The direct model means the contact event data are stored in both the UE and the OE where OE can opt to transmit the data to the FE, which can reduce the use of compute resource on the OE.

From another perspective, UEs and OEs in Fig. \ref{fig:modes} can be operated in a lightweight or heavyweight fashion, depending on how operations are performed. A lightweight UE may run on a mobile device equipped with sensing devices has the minimum duties. A lightweight OE involves no processing and storage of data and only transmits data to FE or to UE.

\begin{figure}[ht]
\centering
\includegraphics[width=0.85\linewidth]{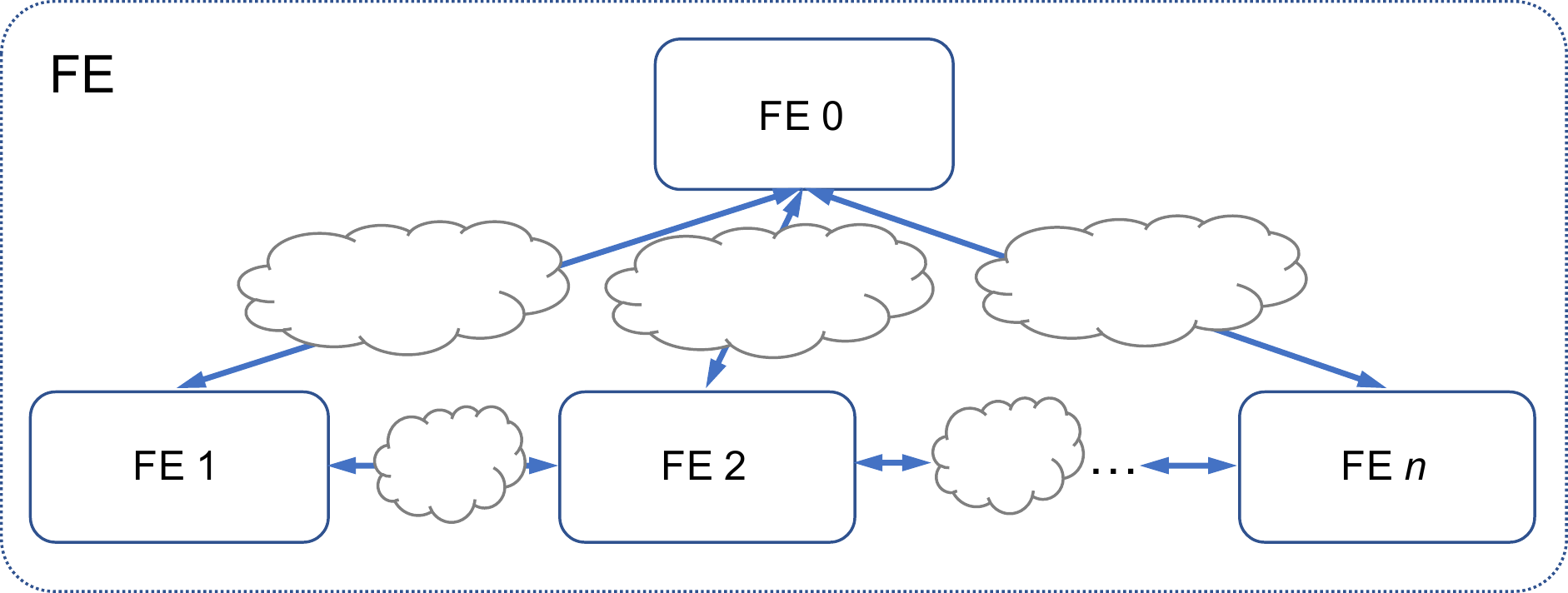}
\caption{Illustration of a virtual FE consisting of multiple FEs}
\label{fig:FE}
\end{figure}

In addition to a single physical FE, it is possible we use multiple FEs as a virtual FE which is shown in Fig. \ref{fig:FE}, where a root FE 0 and $n$ leaf FEs can be interconnected. In a real-world setting, the FE 0 can be an entity of the public health authority and leaf FEs can be infrastructural entities in public facilities handling geographically distributed UEs and OEs. 

\subsection{Discussion on Use Cases}
An IoT-CTS based on the proposed architecture can help access the risk of exposure to contaminated objects or infectious users over time. We basically need two data sources for enabling such a use case. One data source is about the locations of objects and persons which a user has contacted or has been in the proximity of. The other data source is the persons or objects that have been identified as an S/I state where such a data source may be maintained by public health authorities and allow real-time access from authorized users for assisting research and self-evaluation of infection risks. The data sources may be physically stored in multiple repositories such as the FE entities shown in Fig. \ref{fig:modes}. For example, the individual tracing data may be partially stored on a personal device, and partially stored in a data centre or on a content distribution network (CDN) on top of interconnected FEs. The tracing data (who made contacts at which time) on objects or facilities may be stored in its own repositories. In this sense, a mechanism of integrating the repositories to provide seamless data access to users in real-time should be available. In addition, the IoT-CTS can also make responsive data collection for public health agencies. For example, the personal data from facilities with user tracing data can be fused into one data source with other data sources such as clinical data which can be used for studying the transmission and epidemiological patterns, and for determining measures for public health purposes.
 
\section{Analysis and Evaluation}
\subsection{Epidemiological Model of an IoT-CTS}
The proposed solution can basically intervene in the transmissions of an infectious disease at least in the following factors:
\begin{itemize}
    \item Time required to transition from $S$ to $I$ is shortened. This is due to the fact that individual risk can be assessed with medical tests is strengthened. This has been important in the COVID-19 pandemic across countries. 
    \item The individualized strategies can be made when the assisted assessment of the exposure to the potential risks is available. This can result in actions such as reduced interactions between people.
\end{itemize}

The enhanced capability of contact tracing based on the proposed IoT-CTS solution can affect the dynamics of an epidemic, which can be generally explained with the typical SIR model in (\ref{eqn:SIR}). Each individual of the population $N$ can move in three states over time from the susceptible state $S$, infected state $I$, to recovered state $R$. In a time instant $t$, the number of individuals in these states are $S(t)$, $I(t)$, and $R(t)$, respectively, where $S(t)+I(t)+R(t)=N$.

\begin{align}\label{eqn:SIR}
\begin{split}
\frac{dS(t)}{dt} &= - \beta \frac{S(t) I(t)}{N}\\
\frac{dI(t)}{dt} &= \beta \frac{S(t)I(t)}{N} - \gamma I(t)\\
\frac{dR(t)}{dt} &= \gamma I(t)
\end{split}
\end{align}

In (\ref{eqn:SIR}), $\gamma$ is the recovery rate and $\beta$ is the effective contact rate. The typical reproduction number $R_0 = \frac{\beta}{\gamma}$ gets reduced when $\beta$ is reduced and $\gamma$ is kept the same, which can also be explained in (\ref{eqn:R0}) \cite{Jones2007}, where $\tau$ is the probability of infection given contact between an $S$ person and an $I$ person, $\bar{c}$ is the average rate of contact between $S$ and $I$ persons, and $d$ is the duration of infectiousness. 

\begin{equation}\label{eqn:R0}
R_0 = \frac{\beta}{\gamma} = \frac{\tau \cdot \bar{c}}{d}
\end{equation}

We can see that with invariable $\tau$ and $d$, the enhanced contact tracing can reduce the value of $\bar{c}$, which reduces $R_0$ accordingly.


\subsection{Performance Analysis}
Although we can see the overall effectiveness with the proposed IoT-CTS architecture, we need to be able to examine the dynamics with simulations where additional real-world factors such as age, mobility, risk, and hospital capability are to be considered. The Python simulations are made based on the coronavirus simulator (https://tinyurl.com/vhp6pou), where the simulated individuals are initially distributed across the 1$\times$1 unit square plane. Each person randomly moves at the average speed of 0.0042 unit/s which is mapped to the average 2.5 m/s walking speed in the map (with the width of around 600 m) shown in Fig. \ref{tor_example}. Some simulation parameters are shown in Table \ref{Tbl:simuparam} based on the real data from the available information about COVID-19 and public health. The number of hospital beds per 1000 people is the average across OECD countries \cite{oecd2019report}. An infected person is assumed to have less risk when there are enough hospital beds for treatment, and that risk is assumed to be doubled when the hospital beds are insufficient.

\begin{table}[htbp]
\caption{Simulation Parameters}
\label{Tbl:simuparam}
\centering
\begin{tabular}{|c|c||c|c|}
\hline
\textbf{Parameter} & \textbf{Value} & \textbf{Parameter} & \textbf{Value}\\ 
\hline
Population $N$  & 5000 & Average speed (unit/s) & 0.0042 \\
\hline
Mortality rate & 3.4\%  & Hospital beds per 1000 ppl. & 4.7\\
\hline
Infection range & 0.01 & Prob. of infection in range  & 0.03 \\
\hline
\end{tabular}
\end{table}

An IoT-CTS solution based on the proposed architecture can make interventions on a fine-grained individual level. It can at least help reduce the interaction of users as discussed in the previous section and affect two parameters in the simulation, i.e., mobility speed and the probability of infection in range, which is set to 0.0042 unit/s and 5\% by default. The results are compared against the baseline scenario with no measures made.

\begin{figure}[ht]
\centering
\includegraphics[width=0.76\linewidth]{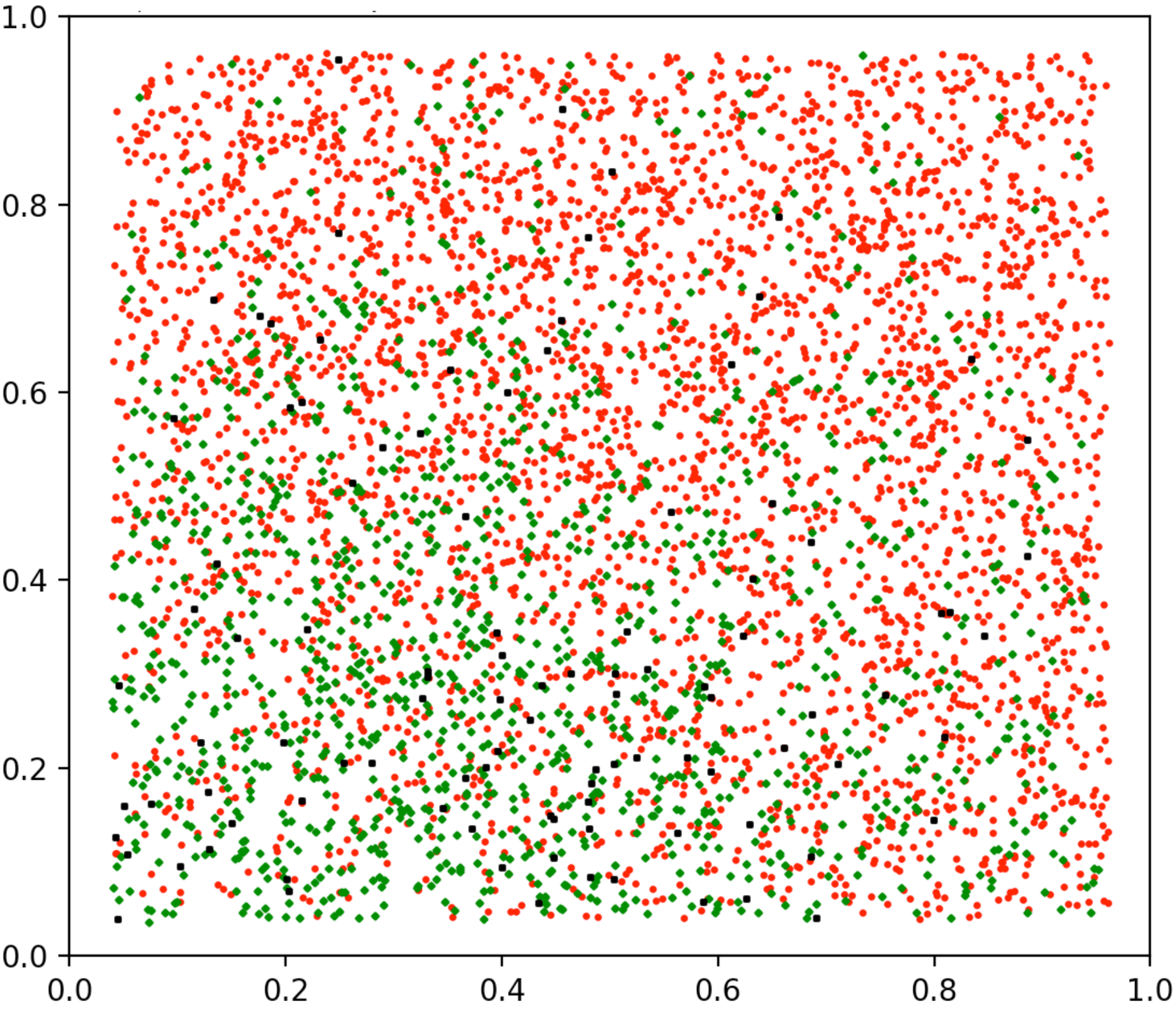}
\caption{Illustration of the running simulation. The red circles, green diamonds, and black squares represent the infectious, recovered, and fatality states of individuals, respectively. }
\label{fig:demo1}
\end{figure}

\begin{figure}[ht]
\centering
\includegraphics[width=0.85\linewidth]{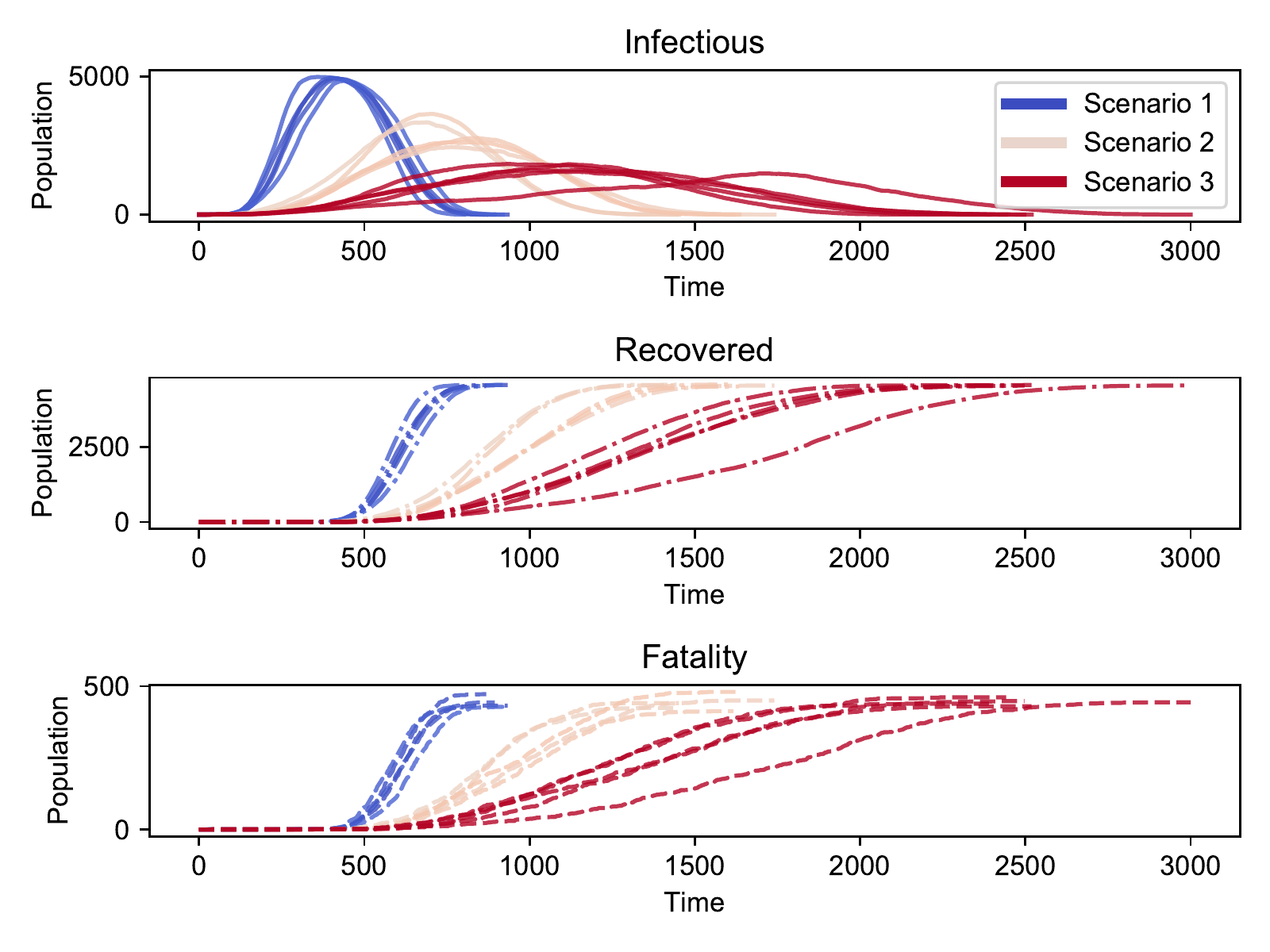}
\caption{Simulation results of three scenarios}
\label{fig:result1}
\end{figure}

\begin{figure}[ht]
\centering
\includegraphics[width=0.86\linewidth]{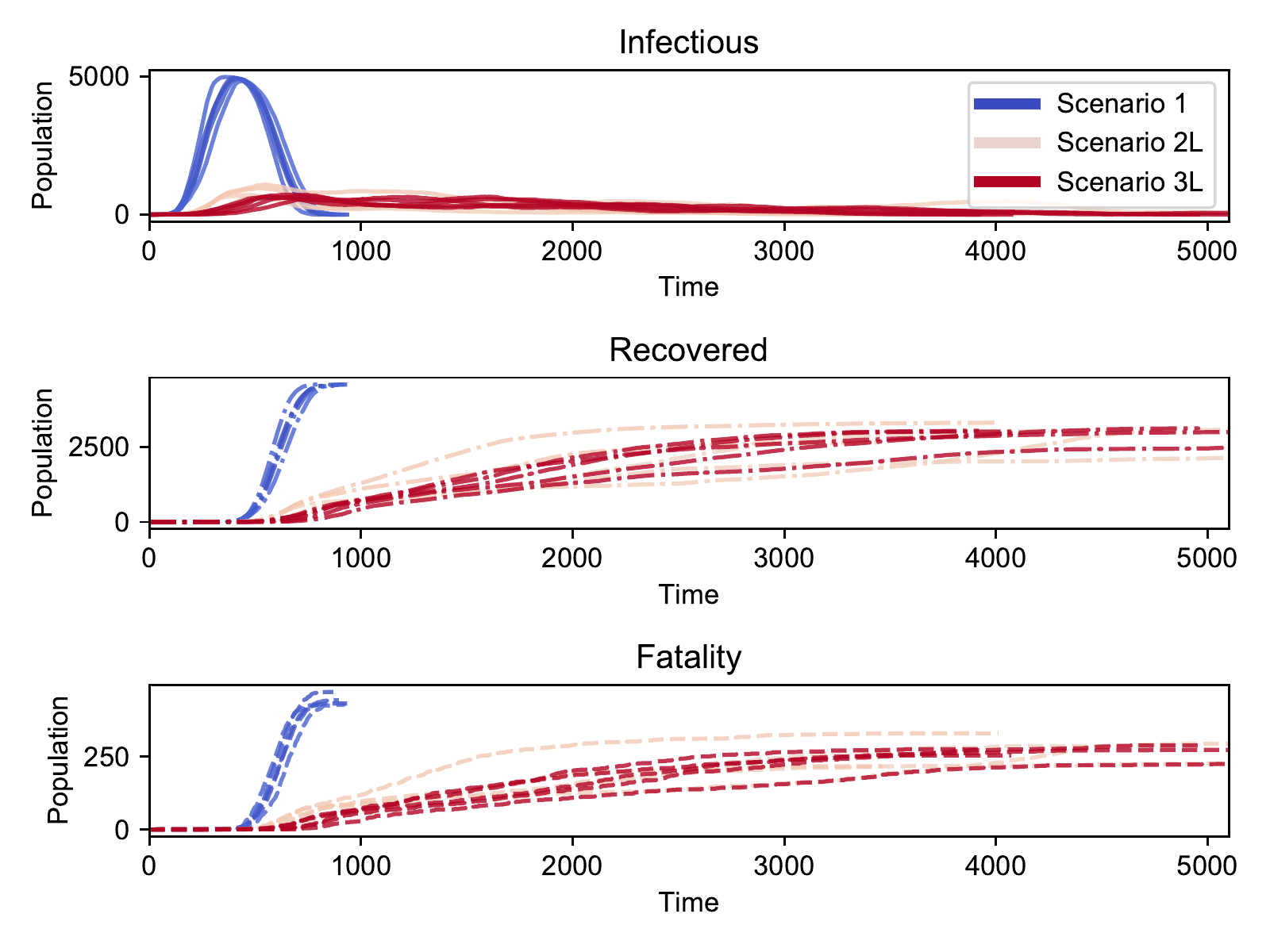}
\caption{Simulation results of three scenarios with a lockdown measure}
\label{fig:result2}
\end{figure}

Three scenarios are evaluated in the simulations. Scenario 1 is the baseline case where no measures are taken and the other scenarios are the cases employing an IoT-CTS. In Scenario 1, the infection chance is 5\% and the average speed of individuals is 0.0042 unit/s. In Scenarios 2, the infection chance is reduced to 2\% and the average speed to 0.002. Scenario 3 is similar to Scenario 2 but the average speed is reduced to 0.001 where the infection chance is kept the same. Scenarios 2 and 3 are used as a result of the fact that people are more aware of the risks and taken actions. Other measures may be able to further reduce the key parameters based on Scenarios 2 and 3, but the scenarios are used to show the essential results based on IoT-CTS solutions. For each scenario, 5 simulation runs are performed.

The intermediate result at time tick 640 for Scenario I is visualized in Fig. \ref{fig:demo1}, where the infectious, recovered, and fatality states of individuals are shown after the first infectious case starts at the time tick 53 at a random location. The statistical results for all three scenarios are shown in Fig. \ref{fig:result1}. We can see overall Scenario 3 has the best results where the number of infectious cases and fatality cases is the lowest of the 3 scenarios. Also, in Scenario 3, while the number of recovered cases is the same as the other two scenarios, its infectious, recovered, and fatality cases take the longest time span. For example, the last infectious case of Scenario 3 occurs at the time tick 3005, where the last recovered case and fatality case occur at a similar time, while these cases of Scenario 1 occur around the time tick 937. Scenario 2 has a better result where the last infectious case occurs at the time tick 1744 but the fatality case occurs at the time tick 1742 which lasts longer. Scenario 2 has a better result than that of Scenario 1, where its number of infectious and fatality cases is lower than those in Scenario 1. We can also see the curves of the infectious cases in Scenarios 2 and 3 are flatter than those of Scenario 1. The similar fatalities among the scenarios are due to the limited hospital resources we have configured for the simulation. In summary, the simulation results essentially show the overall effectiveness of the IoT-CTS solutions based on the proposed architecture.

In Fig. \ref{fig:result2}, the results of the Scenarios 2 and 3 with the use of lockdown measures (referred to as Scenarios 2L and 3L, respectively) are shown, where the lockdown measure for all people is taken when 10\% population is infected. Assuming 92\% population is in compliance with the measure, we can see the number of infectious cases here is much lower than that in Fig. \ref{fig:result1}. Scenario 3L has a better performance than Scenario 2L in the infectious and recovered cases. The results show even in the presence of a strong lockdown measure an IoT-CTS can still play a key role in early containment of the spreads of a disease.


\section{Conclusion}
With the intensive efforts on combating COVID-19, one question we would raise is ``what would we do differently with the latest IoT technologies when facing such an infectious disease?'' Technological advancements in communications, sensing, and computation have enabled powerful tools for infectious disease surveillance and detection in various aspects. Recent advancements in IoT, Tactile Internet and AI allow us to propose novel solutions based on new computation models with ubiquitous and ever-enhanced sensing and networking. It is of utmost importance to start exploring the key fundamentals of the IoT-CTS solutions for the current pandemic and beyond.

The proposed work allows us to design IoT-CTS solutions in a standardized way. With the IoT-CTS architecture, the fine-grained capability of contact tracing can bee utilized in various applications, such as containment of infectious diseases, assisted self-evaluation of infections, and responsive data sharing for collaborative research efforts. The future work includes the improvement of the system architecture and its theoretical models extended from this work.

\ifCLASSOPTIONcaptionsoff
  \newpage
\fi



\bibliographystyle{IEEEtran}

\bibliography{IEEEabrv,./main.bib}

%



%




\end{document}